\documentclass[twocolumn,showpacs,preprintnumbers,amsmath,amssymb,prb,superscriptaddress]{revtex4}

\usepackage{graphicx}
\usepackage{dcolumn}
\usepackage{bm}

\newcommand{\be}{\begin{equation}}
\newcommand{\ee}{\end{equation}}

\newcommand{\beq}{\begin{eqnarray}}
\newcommand{\eeq}{\end{eqnarray}}

\def\eq#1{(\ref{#1})}
\def\H1{\widehat{H}_1}

\begin{document}

\title{Tunneling conductance of a mesoscopic ring with spin-orbit coupling
and  Tomonaga-Luttinger interaction}

\author{M. Pletyukhov}
\affiliation{Institut f\"ur Theoretische Festk\"orperphysik, Universit\"at
Karlsruhe, D-76128 Karlsruhe, Germany}

\author{V. Gritsev}
\affiliation{Department of Physics, Harvard University, Cambridge,
Massachusetts 02138, USA}
\affiliation{D\'epartement de Physique, Universit\'e
de Fribourg, CH-1700 Fribourg, Switzerland}

\author{N. Pauget}
\affiliation{Institut f\"ur Theoretische Festk\"orperphysik, Universit\"at
Karlsruhe, D-76128 Karlsruhe, Germany}

\begin{abstract}
We study the tunneling current
through a mesoscopic two-terminal ring with spin-orbit coupling,
which is threaded by a magnetic flux. The electron-electron interaction
in the ring is described in terms of a Tomonaga-Luttinger model which also
allows us to account for a capacitive coupling between the ring and the gate
electrode. In the regime of weak tunneling, we describe how, at temperatures
lower than the mean level spacing, the peak positions of the conductance
depend on magnetic flux, spin-orbit coupling strength, gate voltage,
charging energy, and interaction parameters (charge and spin velocity and
stiffness).

\end{abstract}



\maketitle

\section{Introduction}
Mesoscopic rings represent an important tool for experimental and
theoretical studies of various phenomena which take place on a
submicrometer scale. The ring geometry allows one to probe
many interesting theoretical predictions. One of the most exciting
phenomena is the generation of geometric phases
which are manifested in the interference patterns of wave packets propagating
in the ring. Along with the well-known Aharonov-Bohm (AB) effect \cite{AB}
which takes place for both spinless and spinful particles, the generation
of a spin-dependent phase is also possible. This effect, sometimes
called the Aharonov-Casher (AC) effect\cite{AC}, may occur in the
transport of electrons when they are subject to sufficiently strong
spin-orbit (SO) coupling. The recent fabrication of HgTe rings
\cite{HgTe} made it possible to directly observe the AC phase.
In earlier experiments with other compounds\cite{Morp,KSN,BKSN}
the signatures of this effect have been also detected.

In order to probe the AC phase it is necessary to have
a tool for manipulating the strength of the spin-orbit coupling.
This is provided by the gate-voltage dependence\cite{nitta} of the Rashba
SO coupling\cite{BR}, which serves as a basis for a construction of a
spin field-effect transistor\cite{DDT}. Changing the magnetotransport
properties of the ring in this way, the
experimentalists are now able to study the AC effect\cite{HgTe,Morp}.

Usually the current through a mesoscopic noninteracting ballistic
ring is described theoretically by means of the Landauer-B\"{u}ttiker
scattering matrix theory\cite{butt}. Geometric phases arising due to both
magnetic flux and SO coupling can be naturally incorporated in this
formalism \cite{Stern,aronov,NMT,FR,M}. Effects of electron-electron
interaction and charging energy are not taken into account in such a
consideration. However, they might be important, for example, in small
quasi-one-dimensional
(quasi-1D) rings or in arrays of such rings fabricated in very recent
experiments\cite{Morp,KSN,BKSN}.

\begin{figure}[t]
\includegraphics[width=8cm,angle=0]{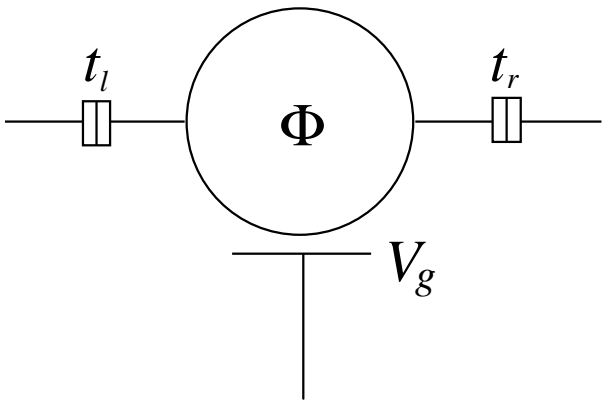}
\caption{The ring threaded by a
magnetic flux $\Phi$ is weakly coupled to the leads
through the tunneling barriers $t_{l}$ and $t_{r}$ and 
capacitively coupled to the gate electrode ($V_g$). }
\label{ringscheme}
\end{figure}

In the present paper we calculate the linear tunneling conductance
of the quasi-1D two-terminal ballistic ring with Rashba SO coupling
threaded by a  magnetic flux. The setup is schematically shown in
Fig.~\ref{ringscheme}. The spectrum of electrons in the ring is
SO-split into two subbands. We will assume electron densities at
which only the lowest radial band is occupied. The electron-electron
interaction inside the ring is modeled by the parameters of the
Tomonaga-Luttinger liquid (TLL), the leads being noninteracting. Assuming
a weak tunneling between the leads and the ring, we compute the
leading term of the Kubo conductance perturbatively expanded in a
series of tunneling elements. We mostly follow the approach of
Ref.~\onlinecite{kin} where a similar problem for spinless fermions was
considered.  We also make use of the bosonization in order to calculate the 
required TLL correlation functions. However, instead of the Matsubara 
formalism, we apply the Keldysh real-time approach to this quasiequilibrium 
problem (cf. Ref.~\onlinecite{meir}). 
Such a combination of the Keldysh technique and bosonization appears
more efficient for a derivation of asymptotic results at
temperatures lower than the mean level spacing of the ring's
spectrum.

After Ref. \onlinecite{kane} it is known that an electron-electron interaction 
strongly renormalizes the height of tunneling barriers between the leads and 
TLL, and therefore at $T=0$ electron transport is suppressed. 
At finite temperatures $T \neq 0$ the linear conductance vanishes as a power 
law of $T$, while the effective width of a conductance peak grows with 
$T \to 0$. In order to ensure the validity of the weak-tunneling 
approximation, in our studies we assume a temperature range where 
the renormalized tunneling rates are smaller than the temperature, 
$\widetilde{\Gamma}_{l,r} \ll T$. 
On the other hand, finite-size effects remain important 
at $T \ll \omega_0$, the single-particle level spacing near the Fermi level.

In the temperature regime $\widetilde{\Gamma}_{l,r} \ll T \ll \omega_0$ 
the linear conductance is represented by a 
sequence of resonance peaks when plotted as a function of gate voltage 
and/or magnetic flux. In our paper we focus on the problem of how the 
distribution
of the conductance peaks depends on the external parameters (magnetic
flux, SO coupling, gate voltage, charging energy) as well as on the
parameters of the Tomonaga-Luttinger interaction. The perturbative
expansion of the linear conductance in tunneling elements is known to break 
down in the  resonance positions. Finding the poles of the
leading term we can establish where the conductance peaks are centered. 
Thus, the study of electron transport in the TLL ring
provides an effective tool of spectroscopy of its many-body states.
Conceptually this is analogous to the study of the tunneling
conductance between the two parallel quantum wires\cite{ZG} which
has been realized experimentally\cite{aus}. We note that a
description of a shape of a particular peak is, however, a different problem
which is usually tackled in a somewhat different manner (cf., e.g.,
Refs.~\onlinecite{fur,fur2,nazarov}), and it will not be addressed
in the present context.

In our paper we extensively discuss the importance of the so-called Klein
factors and zero modes (topological excitations) of the bosonized
Hamiltonian\cite{delft} for the  description of distribution of 
conductance peaks. An accurate account of the
Klein factors is necessary due to the presence of spin-orbit
coupling in the system. The zero-mode sector of TLL decouples from its
``continuous'' (bosonic) sector and contains the whole dependence on external
parameters\cite{kin,faz}. The latter appear in the topological sector
after imposing boundary conditions. We elaborate on the procedure of averaging
the conductance over zero modes in the presence of spin-orbit coupling and 
obtain analytically asymptotic results for the peak positions at
temperatures lower than the mean level spacing. We also reexamine
the case of spinless fermions reproducing the result of Ref.~\onlinecite{kin}
and discuss it in further detail.

It is worthwhile to note that the relevance of
the topological modes for a description of mesoscopic phenomena in the
TLL rings has been already appreciated in various contexts,
including studies of persistent\cite{loss} and Josephson
currents\cite{faz} and the study of the AB phase in chiral
Luttinger liquids\cite{loss2}. The structure of the topological sector in the
presence of  SO coupling has been recently discussed as well in applications 
to persistent\cite{mosk,plet} and Josephson\cite{KKSJ} currents.

The paper is organized as follows. In Sec.~\ref{disper} we briefly
outline the construction of the spectrum of the ring with SO
coupling. In Sec. III we summarize the results emerging from an application
of the Landauer-B\"{u}ttiker formalism to this system. They will be further
used as a reference in the noninteracting limit. In Sec. IV we
present a derivation of the Kubo formula in the real-time approach.
Briefly reviewing the bosonization formalism in Sec. V, we derive an
expression for the dc conductance to be averaged over zero modes.
The procedure of averaging is performed in Sec. VI. We discuss the interplay
of the externally tuned and interaction parameters in the distribution
of the conductance peaks, especially focusing on the modification
of  the Coulomb blockade due to SO coupling.

\section{Mesoscopic rings with Rashba coupling: dispersion relations}
\label{disper}

The two-dimensional electron gas with Rashba spin-orbit coupling is described 
by the Hamiltonian
\be
H = \frac{1}{2 m^*} (p_x^2 +p_y^2) + \alpha_R (\sigma_x p_y - \sigma_y p_x) +
V (r) ,
\label{schrod}
\ee
where $r=\sqrt{x^2+y^2}$. The magnetic field ${\bf B}$ is introduced 
in the kinetic momentum
${\bf p} \to {\bf p} + \frac{e}{c} {\bf A}$ via the gauge potential
${\bf A} = \frac{B}{2} (-y,x,0)$. The radial potential $V (r)$
confining an electron to the ring geometry can be modeled, for
example, either by singular isotropic harmonic oscillator or by
concentric hard walls \cite{plet}. For these or any other types of
the radial confinement the resulting  quasi-one-dimensional spectrum
$\varepsilon_{n \sigma} (k)$ is labeled by the radial band index $n =
0,1,\ldots$, by the angular momentum $\hbar k = \ldots , -\hbar , 0,
\hbar, \ldots$, and by the subband index (chirality) $\sigma =\pm$.
From now on we will put $\hbar=1$.

\begin{figure}[b]
\includegraphics[width=8cm,angle=0]{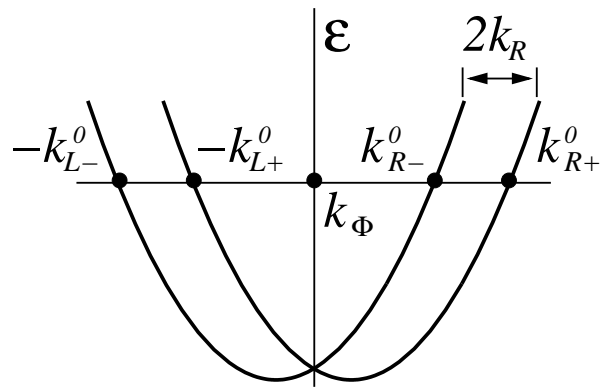}
\caption{The lowest radial band of the quasi-1D mesoscopic
ring SO-split into two subbands.}
\label{sospec}
\end{figure}

If the effective ring's width is much smaller than the ring's
radius, we can neglect the hybridization between the radial bands.
We also assume electron densities at which only the lowest radial
band ($n=0$) is occupied. Thus, we effectively consider the strictly
one-dimensional spectrum (see Fig.~\ref{sospec}) which has a
parabolic shape and is SO-split into two subbands\cite{plet}:
\be
\varepsilon_{\sigma} (k) \equiv \varepsilon_{0 \sigma} (k) 
= \frac{2 \pi^2}{m^* L^2} (k - k_{\Phi} -\sigma k_R)^2.
\label{spectrum}
\ee
Here $L$ is the ring's perimeter, $k_{\Phi} = \Phi/\Phi_0$
is a number of flux quanta $\Phi_0$ threading the ring, and  the parameter
\be
k_R = \sqrt{\frac14+\left(\frac{\alpha_R m^{*} L}{2 \pi}\right)^2} - \frac12
\ee
depends on the Rashba coupling $\alpha_R$.

Linearizing the spectrum \eq{spectrum} near the Fermi energy, we obtain the 
four branches
\be
\varepsilon_{\eta \sigma} (k) = \omega_0 (k - k_{\eta \sigma}^0) \equiv 
\omega_0 (k- \eta k_F - k_{\Phi} - \sigma k_R),
\label{linspec}
\ee
specified by $\eta = \pm$ (or $\eta =R,L$) and $\sigma = \pm$. The
Fermi angular velocity $\omega_0  = \left( \frac{2 \pi}{L}\right)^2
\frac{k_F}{m^*}$ defines the level spacing of the spectrum
\eq{linspec}, and $k_F$ is the Fermi angular momentum in absence of a
magnetic field and SO coupling.

\section{Conductance of the mesoscopic ring: noninteracting electrons}

Let us consider the conductance of the ring attached to the
semi-infinite leads (Fig.~\ref{ringscheme}). For noninteracting electrons
it can be easily found in the framework of the scattering matrix
theory\cite{butt}.

It is instructive to consider first the case of spinless fermions with the
two linearization points $k_{R/L}^0$. One finds that in the zero-temperature
limit and for the angle $\pi$ between the junctions to the leads the dc 
conductance reads
\cite{butt}
\begin{widetext}
\be
G (k_F , k_{\Phi}) = \frac{e^2}{2 \pi} \,\, \frac{16 \epsilon_l \epsilon_r 
\sin^2 k_F
\pi \cos^2 k_{\Phi} \pi}{[-2 \alpha_l \alpha_r + (1+ \gamma_l \gamma_r) 
\cos 2 \pi k_F -
2 \beta_l \beta_r \cos 2 \pi k_{\Phi}]^2 + (1- \gamma_l \gamma_r)^2 
\sin^2 2 \pi k_F},
\label{noncond}
\ee
\end{widetext}
where $\epsilon_{l/r}$, $\gamma_{l/r} = - \sqrt{1 - 2
\epsilon_{l/r}}$, $\alpha_{l/r} = - \frac12 (1 + \gamma_{l/r} )$,
and $\beta_{l/r} = \frac12 (1 -\gamma_{l/r})$ are  the
phenomenological parameters describing scattering in a T-shaped
(left $l$ or right $r$) junction. The number of flux quanta is given
by  $k_{\Phi}= \frac12 (k_R^0 - k_L^0)$, while the quantity $k_F =
\frac12 (k_R^0+k_L^0)$ corresponds to the Fermi momentum at zero
flux. It can be replaced by $k_F \to N_0 + \frac{\Delta
\mu}{\omega_0}$, where $\Delta \mu$ is a difference between the
chemical potential of the leads and the Fermi energy of the ring,
and the integer $N_0$ is related to the number $2 N_0+1$ of
electrons in the ring at $\Delta \mu =0$. Since the expression
\eq{noncond} is periodic in $k_F$, the integer part of $k_F$ can be
discarded. Thus, the conductance \eq{noncond} actually depends on
the fractional part of $ \frac{\Delta \mu}{\omega_0}$. For
future references we introduce the parameter $k_{\mu} = \frac{\Delta
\mu}{\omega_0} -\frac12$.

\begin{figure}[b]
\includegraphics[width=8cm,angle=0]{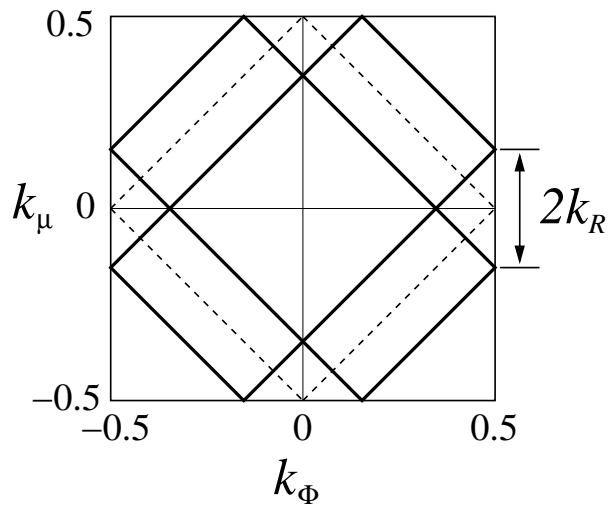}
\caption{Splitting of the conductance peaks (solid lines) due
to SO coupling. The dashed lines correspond to $k_R =0$.}
\label{condpeakLB}
\end{figure}

In the weak-tunneling limit $\epsilon_{l/r} \ll 1$ the conductance
\eq{noncond} approximately equals
\be
G \approx \frac{e^2}{2 \pi} \,\, \frac{4 \epsilon_l \epsilon_r \sin^2 k_F
\pi \cos^2 k_{\Phi} \pi}{(\cos 2 \pi k_F - \cos 2 \pi k_{\Phi})^2 + \frac14
(\epsilon_l + \epsilon_r)^2 \sin^2 2 \pi k_F}.
\label{lowt}
\ee
As a function $k_{\mu}$ and $k_{\Phi}$, it represents a sequence of
Breit-Wigner resonances. The conductance peaks occur when the
resonance condition $\cos 2 \pi k_F = \cos 2 \pi k_{\Phi}$ is
fulfilled --- i.e., at the values of the parameters
\be
k_F + k_{\Phi} = n_{R}, \quad k_F - k_{\Phi} = n_{L},
\label{res0}
\ee
where $n_R$ and $n_L$ are arbitrary integers. We note that in the 
weak-tunneling limit the resonance condition \eq{res0} remains valid for 
arbitrary  angle $x_l -x_r$ between the junctions, while the  shape of  
Breit-Wigner resonances is quite sensitive to the value of $x_l -x_r$.

It has been demonstrated in Ref.~\onlinecite{aronov} that for electrons with
nonzero SO coupling and negligible Zeeman splitting the conductance of the
mesoscopic
ring is given by the sum of the two contributions: $G (k_F, k_{\Phi}+k_R)$ and
$G (k_F, k_{\Phi}-k_R)$. In other words, the net effect of the SO coupling
for noninteracting electrons is the generation of the different effective
flux values for the  different channels. Therefore, the pattern of the
conductance maxima at $k_R \neq 0$ is determined by the resonance conditions
\be
k_F + (k_{\Phi} \pm k_R) = n_{R\pm}, \quad k_F - (k_{\Phi} \pm k_R) = n_{L\pm},
\ee
where $n_{R \pm}$ and $n_{L \pm}$ are arbitrary integers. Recalling
that effectively $k_{\mu} = k_F - \frac12$, we show in
Fig.~\ref{condpeakLB} how the arrangement of the conductance peaks
is modified by SO coupling.

\section{Kubo formula}

In order to take into account  effects of the electron-electron
interaction on the distribution of conductance peaks, we discuss in
this section the Kubo formula for the linear conductance. Although
this expression is very standard, we rederive it using the Keldysh
formalism. In doing this, we pursue two objectives. First, we would
like to have better control of the approximations used (similar to
those made in Ref.~\onlinecite{kin}). Second, we would like to
deduce an expression for the conductance in a real-time
representation. Its advantage for the ring geometry will be
discussed in the next section where the calculation of
time-dependent finite-size TLL correlation functions is concerned.

In the second-quantized formulation the mesoscopic ring attached
to the leads is described by the Hamiltonian
\be
H = H_l + H_r + H_c + H_T.
\ee
The left/right lead is described by a Fermi-liquid
Hamiltonian $H_{l/r} = \int dx c_{l/r}^{\dagger} (x) \left(
\frac{p^2}{2m^*} - \mu \right) c_{l/r} (x)$, and the tunneling term
is $H_T = \sum_{l,r} [ t_{l/r} c_{l/r}^{\dagger} (x_{l/r}) \psi
(x_{l/r})+ {\rm H.c.}]$. Here $c_{l/r}$ and $\psi$ are the field 
operators in the leads and in the ring, respectively. 

The Hamiltonian of the central part (ring) $H_c [\psi^{\dagger},
\psi]$ can have any interaction term in addition to the kinetic term. 
In our consideration we will model the electron-electron interaction in the 
ring by the Tomonaga-Luttinger liquid which includes only forward-scattering
processes (``density-density''-type interaction). In the framework
of this model it is also possible to take into account the charging effects. 
They originate from a capacitive coupling of the ring to the gate electrode 
and are described by the Hamiltonian $E_c (\hat{N}_{ring} - \frac{1}{e} C_g
V_g)^2$, with the charging energy $E_c = e^2/2 C_g$. Here $C_g$ is the gate 
capacitance and $\hat{N}_{ring}$ is the number operator of electrons in the 
ring.

The linear response of the system to an applied time-dependent bias
voltage is described by the Kubo formula for the ac conductance
\cite{mahan}:
\be
G (\Omega) = - \frac{1}{\Omega} \int_{-\infty}^{t} d t' e^{- i \Omega ( t'-t)} 
\langle [\hat{I}_l (t), \hat{I}_r (t') ]\rangle ,
\label{kubo}
\ee
where $\hat{I}_{l/r} (t) = i e [t_{l/r} \hat{c}_{l/r}^{\dagger} (x_{l/r}, t) 
\hat{\psi} (x_{l/r},t)- {\rm H.c.} ]$ is a current operator in the left/right 
junction written in the Heisenberg representation.

In the weak-tunneling limit the expression \eq{kubo} can be expanded in a 
series of $H_T$. We make use of the real-time Keldysh diagrammatic technique, 
and for $t>t'$ we replace $\langle [\hat{I}_l (t), \hat{I}_r (t') ]\rangle$ by
\be
\langle T_t \hat{I}_l (t) \hat{I}_r (t') \rangle - \langle \tilde{T}_t 
\hat{I}_l (t) \hat{I}_r (t') \rangle \equiv 2 i {\rm Im} \langle T_t \hat{I}_l 
(t) \hat{I}_r (t') \rangle.
\label{iicorr}
\ee
When expressed on the Keldysh contour, it reads
\begin{widetext}
\be
\langle T_t \hat{I}_l (t), \hat{I}_r (t') \rangle =\langle 
\tilde{T}_t e^{i \int_{-\infty}^t H_T (t'') dt''} I_l (t) 
T_t e^{-i \int_{-\infty}^t H_T (t'') dt''}
\tilde{T}_t e^{i \int_{-\infty}^{t''} H_T (t'') dt''} I_r (t') 
T_t e^{-i \int_{-\infty}^{t'} H_T (t'') dt''}\rangle ,
\ee
\end{widetext}
where the operators without carets refer to the interaction ($H_T$)
representation.

Expanding \eq{iicorr} to the second order in $H_T$, we obtain
\beq
& & \int_{-\infty}^{t} dt_1 \int_{-\infty}^{t'} dt'_1 
\langle [[ I_l (t), H_T (t_1)] , [H_T (t'_1), I_r (t')]] \rangle \nonumber \\
&-& \int_{-\infty}^{t} dt_1 \int_{-\infty}^{t_1} dt_2 
\langle [[[ I_l (t), H_T (t_1)] , H_T (t_2)], I_r (t')] \rangle \nonumber \\
&-&  \int_{-\infty}^{t'} dt'_1 \int_{-\infty}^{t'_1} dt'_2 
\langle [I_l (t), [ H_T (t'_2), [ H_T (t'_1), I_r (t')]]] \rangle. \nonumber
\eeq

The next step is to perform averaging over the leads' states.
While doing this, we meet the following combinations:
(a) $(G^R_l -G^A_l)(G^R_r -G^A_r)$, (b) $G^K_l(G^R_r -G^A_r)$,
(c) $(G^R_l -G^A_l) G^K_r$, and  (d) $G^K_l G^K_r$. Here $G^{R,A}_{l/r}$
and $G^K_{l/r}$ are the momentum-averaged retarded, advanced, and Keldysh 
functions of the leads in the real-time representation
\beq
(G^{R}- G^{A})_{l/r} (t) = -i \langle \{ c_{l/r} (t), c_{l/r}^{\dagger} (0)\}
\rangle = - 2 \pi i \delta (t) \frac{\nu_{l/r}}{V_{l/r}}, \nonumber \\
G^{K}_{l/r} (t) =  -i \langle [ c_{l/r} (t), c_{l/r}^{\dagger} (0)]\rangle
= -\frac{2 \pi}{\beta \sinh (\pi t/\beta)} \frac{\nu_{l/r}}{V_{l/r}},
\nonumber
\eeq
and $\nu_{l/r}$ is the density of states in the left/right 
lead at the Fermi level.

One can straightforwardly  prove that the combinations (a) and (b) vanish
identically. The combination (c) gives the following contribution to
the conductance:
\beq
G^{(c)} (\Omega)  =  e^2 \Gamma_l \Gamma_r L^2 \int_{-\infty}^{0}
\int_{-\infty}^0 d t_1  d t_2 \frac{e^{-i \Omega t_1} (1-e^{-i \Omega t_2})}{2 
i \Omega \beta \sinh [\pi t_2 /\beta]}
\nonumber \\
\times  {\rm Re} \langle \{[[\psi_l (0), \psi^{\dagger}_l (0)], \psi_r (t_1)], 
\psi_r^{\dagger} (t_1+t_2)\}\rangle, \quad \label{condc}
\eeq
where $\Gamma_{l/r} = 2 \pi \nu_{l/r} |t_{l/r}|^2/(V_{l/r} L)$ and
$V_{l/r}$ is the volume of the left/right lead.

From Eq. \eq{condc} we derive an expression for the dc conductance ($\Omega=0$)
at zero temperature
\beq
G^{(c)} &=& \frac{e^2}{2 \pi} \Gamma_l \Gamma_r L^2 \int_{-\infty}^{0} 
\int_{-\infty}^0 d t_1  d t_2
\nonumber \\
& \times & {\rm Re} \langle \{[[\psi_l (0), \psi^{\dagger}_l (0)], 
\psi_r (t_1)], \psi_r^{\dagger} (t_1+t_2)\}\rangle .
\label{gsk0}
\eeq
Using the operator identities
\beq
\{ [C, A], B\} + \{ [C, B], A \}
= [C, \{ A, B \} ], \\
\{ C, \{ A , B \} \} - \{ A, \{ C , B \} \} =[[ C,  A ],B],
\eeq
we rewrite Eq. \eq{gsk0},
\beq
G^{(c)} &=& \frac{e^2}{4 \pi} \Gamma_l \Gamma_r L^2 \int_{-\infty}^{0} 
\int_{-\infty}^0 d t_1  d t_2
\nonumber \\
& \times & \left\{ {\rm Re} \langle \{[[\psi_l (0), \psi^{\dagger}_l (0)], 
\psi_r (t_1)], \psi_r^{\dagger} (t_2)\}\rangle \right. \nonumber \\
& & \left. + {\rm Re} \langle [[\psi_l (0), \psi^{\dagger}_l (0)], \{\psi_r 
(t_1), \psi_r^{\dagger} (t_2)\}]\rangle \right\},
\label{gsk1}
\eeq
and further express
\beq
& & \langle \{ [[\psi_l (0), \psi_l^{\dagger} (0)], \psi_r (t_1) ], 
\psi_r^{\dagger} (t_2) \} \rangle \nonumber   \\
&=& \langle \{ \{ \psi_l (0), \psi_r^{\dagger} (t_2) \}, \{ \psi_r (t_1) , 
\psi_l^{\dagger} (0) \} \} \rangle \nonumber \\
&+&  \langle [[\psi_r^{\dagger} (t_2), \{ \psi_l^{\dagger} (0), \psi_r (t_1)\} 
], \psi_l (0) ] \rangle \nonumber \\
&-& \langle \{ \{ \psi_l^{\dagger} (0), \{\psi_l (0), \psi_r (t_1)  \}\}, 
\psi_r^{\dagger} (t_2) \} \rangle .
\label{gskexp}
\eeq

It is obvious that in the noninteracting limit the only term
$\langle \{ \{ \psi_l (0), \psi_r^{\dagger} (t_2) \}, \{ \psi_r (t_1) , 
\psi_l^{\dagger} (0) \} \} \rangle$ survives, since the other terms vanish due 
to the fermionic commutation relations. We approximate the dc conductance in 
the interacting case by this dominant contribution
\beq
G &\approx & \frac{e^2}{4 \pi} \Gamma_l\Gamma_r L^2 \int_{-\infty}^0  
\int_{-\infty}^0 d t_1 d t_2 \nonumber \\
& \times & {\rm Re} \langle \{ \{ \psi_l (0), \psi_r^{\dagger} (t_2) \}, 
\{ \psi_r (t_1) , \psi_l^{\dagger} (0) \} \} \rangle .
\label{gsk2}
\eeq
Splitting the four-particle correlator, one can recover the formula
$G \approx (e^2/2 \pi)  \Gamma_l\Gamma_r L^2 |G^R (\omega=0, x_l
-x_r)|^2$ from Ref.~\onlinecite{kin}, where $G^R (\omega=0, x_l
-x_r)$ is a zero-frequency retarded Green's function for interacting
electrons in the ring. This approximation physically means that one
scattering event is completed before another takes place. In
general, the TLL correlation functions of any order can be
calculated within the bosonization approach, and this approximation
can be relaxed.

The combination (d) with $G^K_l G^K_r$ also  gives a finite contribution
to the conductance, which, however, vanishes in the noninteracting limit
as well. Therefore, we will neglect it on the same ground as  we have just
neglected the subdominant terms in Eqs. \eq{gsk1} and \eq{gskexp}.

\section{Bosonization}

\subsection{Spinless case}

In order to compute the four-particle correlator \eq{gsk2},
we will make use of the bosonization technique\cite{delft}.

Let us consider for simplicity the spinless case. We introduce
the shorthand notations
for the fermionic fields $\psi_l \equiv \psi (x_l)$ and
$\psi_r \equiv \psi (x_r)$,
where $x_{l}$ and $x_{r}$ are the angle coordinates of the
left and right junctions.
In the following we assume that $x_l = 0$ and $x_r =\pi$.

In the bosonization the fields $\psi_{l/r}$ are represented
as a sum of the  right- ($\eta = +$, or $R$)
and left- ($\eta = -$, or $L$) moving components,
\be
\psi_{l/r} = \psi_{l/r,R} + \psi_{l/r,L} =
F_{l/r,R} \psi_{l/r,R}^{b}  + F_{l/r,L} \psi_{l/r,L}^{b},
\ee
and each of $\psi_{l/r,\eta}$ consists of a topological
part $F_{l/r,\eta}$ and a bosonic part $\psi_{l/r,\eta}^{b}$
commuting with each other: $[F, \psi^b]=0$.

The bosonic part is given by
\beq
\psi_{\eta}^b (x) &=& \frac{1}{\sqrt{L \tilde{\alpha}}}
e^{- i \sqrt{2 \pi} \phi_{\eta} (x)}, \\
\phi_{\eta} (x) &=& i \sum_{k=1}^{\infty}
\frac{e^{-\frac12 \tilde{\alpha} k}}{\sqrt{2 \pi k}}
\left( e^{i \eta k x} b_{\eta k}
- e^{-i \eta k x}  b_{\eta k}^{\dagger}\right),
\nonumber
\eeq
where $\tilde{\alpha} = \frac{2 \pi \alpha}{L}$ is a
small dimensionless cutoff parameter and the operators
$b_{\eta k}$, $b_{\eta k}^{\dagger}$ satisfy the bosonic
commutation relation $[b_{\eta k}, b_{\eta' k'}^{\dagger}]
= \delta_{\eta \eta'} \delta_{k k'}$.

The topological part is important for the finite-size TLL
with periodic boundary conditions.
It includes Klein factors $F_{\eta}$, zero-mode operators
$N_{\eta}$, and the linearization points $k_{\eta}^0$
(see Fig.~\ref{sospec} assuming $k_R =0$):
\be
 F_{l/r,\eta} = F_{\eta} e^{i \eta (N_{\eta} - k_{\eta}^0) x_{l/r}}.
\ee
The zero-mode operators $N_{\eta} = N_{\eta}^{\dagger}$
take integer values, and the following relations are satisfied\cite{delft}:
\beq
[F_{\eta}, N_{\eta'}] &=& F_{\eta} \delta_{\eta \eta'},
\label{FNcom} \\
\{F_{\eta}, F_{\eta'}^{\dagger} \} &=& 2 \delta_{\eta \eta'}, \\
\{F_{\eta}, F_{\eta'} \} &=&
\{F_{\eta}^{\dagger}, F_{\eta'}^{\dagger} \} = 0
\quad {\rm for} \quad \eta \neq \eta'.
\eeq

The bosonized TLL Hamiltonian $H_{TLL} \equiv H_c = H_b + H_0$
consists of a ``continuous'' (bosonic) $H_b$ part and a topological
$H_0$ part which are decoupled from each other. Therefore, the
factorization of $\psi_{l/r,\eta}$ into $F_{l/r,\eta}$ and
$\psi_{l/r,\eta}^{b}$ takes place at any time instant:
\be
\psi_{l/r} (t) = F_{l/r,R} (t) \psi_{l/r,R}^{b} (t)
+ F_{l/r,L} (t) \psi_{l/r,L}^{b} (t),
\ee
where the time evolutions of $\psi_{l/r,\eta}^{b} (t)$ and
$F_{l/r,\eta} (t)$ are governed by $H_b$ and $H_0$, respectively.
By the same reason the statistical averagings in both bosonic and
topological sectors are independent of each other.

The bosonic part of the TLL Hamiltonian is given by
\be
H_b = \frac{2 \pi v}{L} \sum_{a=1,2} \sum_{k=1}^{\infty}
k d^{\dagger}_{ak} d_{ak} ,
\label{bosham}
\ee
where $v$ is the so-called charge velocity (the renormalization
of the Fermi velocity $v_0 \equiv \frac{L \omega_0}{2 \pi}$).
The operators $d_{ak}$, $d^{\dagger}_{a k}$ ($a=1,2$) are obtained
from $b_{\eta k}$, $b_{\eta k}^{\dagger}$ by the canonical
transformation \eq{bogtr}. The latter
depends on the interaction parameter $\gamma  = \frac12 (\frac{1}{K} +K)$,
where $K$ is the so-called charge stiffness. For repulsive interactions
$K<1$, while in the noninteracting limit $K=\gamma=1$ and $v=v_0$.

The topological part of the  TLL Hamiltonian is
\be
H_0 = \sum_{\eta} \left(a_0 \tilde{N}_{\eta}^2
+ a_1 \tilde{N}_{\eta} \tilde{N}_{-\eta} \right) ,
\ee
where $a_{0,1}= \frac{\omega_0}{4} (\tilde{\nu} \pm \lambda)$ and
\be\label{nu}
\tilde{\nu} =  \nu + \frac{4 E_c}{\omega_0}, \quad \nu
= \frac{v}{K v_0}, \quad \lambda = \frac{v K}{v_0} .
\ee
The topological numbers $\tilde{N}_{\eta} = N_{\eta} - k_{\eta}$
are shifted by $k_{\eta} = k_{\eta}^0 +\delta k_{\mu}$,
where
\be
\delta k_{\mu} = \frac{4 E_c (\frac{1}{e} C_g V_g - 2 N_0)
+ 2 \Delta \mu - \omega_0}{2 \tilde{\nu} \omega_0}
\ee
redefines the linearization points $k_{\eta}^0$ in order to
include the dependence on $\Delta \mu$ and the
gate voltage $V_g$. In the basis $N= N_R + N_L$, $J =N_R - N_L$,
the Hamiltonian $H_0$ acquires the diagonal form
\be
H_0 = \frac{\omega_0}{4} [\tilde{\nu} \tilde{N}^2 + \lambda \tilde{J}^2],
\ee
where $\tilde{N} = N - 2 k_{\mu}$, $\tilde{J} = J - 2 k_{\Phi}$,
and $k_{\mu}=N_0 + \delta k_{\mu}$. One can observe that the whole
dependence on $\Delta \mu$, $V_g$, and $\Phi$ is included in the
topological sector.

Using the commutation relations \eq{FNcom} we find the time
evolution of the Klein factors
\be
F_{\eta} (t)= e^{i H_0 t} F_{\eta} e^{-i H_0 t}
= F_{\eta} e^{-i t P_{\eta} + i t a_0},
\label{evklein}
\ee
where
\be
P_{\eta} = 2 a_0 \tilde{N}_{\eta} + 2 a_1 \tilde{N}_{-\eta}
= \frac{\omega_0}{2} [\tilde{\nu} \tilde{N} \pm \lambda \tilde{J}].
\ee

The details of the time evolution of the bosonic fields are presented
in  Appendix \ref{bosonic}. In fact, they are not very important for
our purpose. We will only exploit  the fact that the average of the
bosonic fields,
\be
g^b (t; \gamma) = \langle \psi_{lR}^{b} (t) \psi_{rR}^{b \dagger} (0)\rangle
\equiv \langle \psi_{lL}^{b} (t) \psi_{rL}^{b \dagger} (0)\rangle,
\label{fser}
\ee
is a periodic function of time which can be expanded in a Fourier series
\be
g^b (t; \gamma) = \sum_{p=0}^{\infty} g_p (\gamma) e^{-i p \omega t},
\label{fourex}
\ee
with frequency $\omega = \frac{2 \pi v}{L}$ and real-valued
coefficients $g_p (\gamma)$. Note that the summation in Eq. \eq{fourex}
is performed only over non-negative integers.

The real-time periodicity of $g^b (t; \gamma)$ is inherited from the
spatial periodic boundary conditions. The occurrence of the Fourier
series \eq{fourex} allows us to perform all time integrals explicitly.
The analysis of the remaining series is a much simpler task.

Let us make yet another approximation in the spirit of 
Ref.~\onlinecite{kin}. In particular, we split the four-particle
bosonic correlator in \eq{gsk2}, neglecting the anomalous averages
(e.g., $\langle \psi^b \psi^b \rangle$), the left-right mixing
(e.g., $\langle \psi_L^b \psi_R^{b \dagger} \rangle$), and the vertex
corrections (averages of operators at the same spatial point, e.g.,
$\langle \psi_r^b \psi_r^{b \dagger} \rangle$) in the {\it bosonic}
(continuous) sector. At the same time, we do not split the topological part 
of the four-particle correlator (unlike has been done in Ref.~\onlinecite{kin})
and perform a single averaging of the whole over zero modes. 

Implementing this procedure, we obtain
\beq
& & \langle \psi_l (0) \psi_r^{\dagger} (t_2) \psi_r ( t_1 )
\psi_l^{\dagger} (0) \rangle \approx  g^{b*} (t_2)  g^b (t_1) \\
& & \times \sum_{\eta_1 , \eta_2}
\left\langle  F_{l, \eta_2} (0) F_{r, \eta_2}^{\dagger} (t_2)
F_{r, \eta_1} (t_1)  F_{l, \eta_1}^{\dagger} (0) \right\rangle_{z.m.} .
\nonumber
\eeq
where $\langle \cdots \rangle_{z.m.}$ implies averaging over zero modes 
to be discussed later. Collecting all contributions, we find
\beq
&G & \approx \frac{e^2}{2 \pi} \Gamma_l \Gamma_r L^2
\sum_{p_1, p_2 =0}^{\infty} g_{p_1}
(\gamma) g_{p_2} (\gamma) \int_{-\infty}^0 \int_{-\infty}^0 d t_1 d t_2
\nonumber \\
 &\times & \sum_{\eta = \pm} {\rm Re} \left\langle
\left( e^{i t_1 (\omega p_1 + a_0 + P_{\eta})}
- e^{-i t_1 (\omega p_1  + a_0 - P_{\eta})}\right) \right. \nonumber \\
& & \times \left( e^{-i t_2 (\omega p_2 + a_0 + P_{\eta})}
- e^{i t_2 (\omega p_2 + a_0 - P_{\eta})}\right) \label{Gtime}\\
&+& e^{i (N_{\eta}+N_{-\eta}) \pi }
\left(e^{i \eta t_1 ( \omega  p_1 + a_0 +P_{\eta})}
- e^{-i \eta t_1 ( \omega  p_1 + a_0 - P_{\eta})} \right) \nonumber \\
& &  \left. \times\left(e^{-i \eta t_2 ( \omega  p_2 + a_0 + P_{-\eta})}
- e^{i \eta t_2 (\omega  p_2+ a_0 - P_{-\eta})}\right) \right\rangle_{z.m.}.
\nonumber
\eeq
Introducing
\be
A_{\eta}^{\pm} = \sum_{p=0}^{\infty} 
\frac{g_p (\gamma)}{\omega p + a_0 \pm P_{\eta}}
\label{fexp}
\ee
and  $A_{\eta} =  A_{\eta}^+ + A_{\eta}^-$, we can cast Eq.~\eq{Gtime}
into the form
\beq
G & \approx&  \frac{e^2}{2 \pi} \Gamma_l \Gamma_r L^2
\left\langle A_{R}^2 + A_{L}^2  \right. \label{conds0} \\
&+& \left. 2 A_{R} A_{L} \cos (\tilde{N}_R+ \tilde{N}_L +2 \delta k_{\mu})
\pi \right\rangle_{z.m.}. \nonumber
\eeq
We remark that the alteration of the angle $x_l -x_r$ between the junctions 
would only modify the Fourier coefficients $g_p (\gamma)$ in Eq.~\eq{fexp} 
as well as the relative phase of the interference term $\sim A_R A_L$ in 
Eq.~\eq{conds0}. Meanwhile, the poles of $A_{\eta}^{\pm}$ in Eq.~\eq{fexp} 
are not sensitive to the value of $x_l - x_r$.

In order to treat further the expression \eq{conds0} we need to establish 
an efficient procedure of averaging over zero modes. But first we are going 
to discuss  the modification of the conductance \eq{conds0} caused by the 
presence of spin degrees of freedom and by spin-orbit coupling.

\subsection{Spinful case}

Performing a similar bosonization procedure in the spinful case, we
obtain the following expression for the dc conductance
\beq
G & \approx&  \frac{e^2}{2 \pi} \Gamma_l \Gamma_r L^2 \sum_{\sigma = \pm}
\left\langle A_{R\sigma}^2 + A_{L\sigma}^2  \right.  \label{spincond} \\
&+& \left. 2 A_{R\sigma} A_{L\sigma}
\cos (\tilde{N}_{R \sigma}+\tilde{N}_{L \sigma} + 2 \delta k_{\mu})
\pi \right\rangle_{z.m.}. \nonumber
\eeq
The zero-mode operators $N_{\eta\sigma}$ with integer eigenvalues are
shifted to $\tilde{N}_{\eta \sigma} = N_{\eta \sigma} - k_{\eta \sigma}$
by $k_{\eta \sigma} = k_{\eta \sigma}^0 + \delta k_{\mu}$, where
\be
\delta k_{\mu} = \frac{4 E_c (\frac{1}{e} C_g V_g - 4 N_0) + 2 \Delta \mu
- \omega_0}{2 \tilde{\nu}_c \omega_0} .
\ee
The integer $N_0 = \frac14 \sum_{\eta,\sigma} k_{\eta \sigma}^0$
is related to the number $4 N_0 +2$ of electrons in the ring when
the parameters
\beq
k_{\Phi} &=& \frac14 \sum_{\sigma} (k_{R \sigma} - k_{L \sigma}), \\
k_{B,R} &=& \frac14 \sum_{\sigma} \sigma (k_{R\sigma} \pm k_{L\sigma})
\eeq
equal zero. The parameter $k_B$ vanishes in the absence of a Zeeman
interaction. The parameter
\be
k_{\mu} = \frac14 \sum_{\eta, \sigma} k_{\eta \sigma}
= N_0 + \delta k_{\mu}
\ee
contains the dependence on $\Delta \mu$ and $V_g$.

Like in the spinless case, it is convenient to introduce
\be
\tilde{\nu}_c =  \nu_c + \frac{8 E_c}{\omega_0}, \quad \nu_{c,s} =
\frac{v_{c,s}}{K_{c,s} v_0},
\quad \lambda_{c,s} = \frac{v_{c,s} K_{c,s}}{v_0},
\label{vcspin}
\ee
and  $\omega_{c,s} = \frac{2 \pi v_{c,s}}{L}$, and $\gamma_{c,s} =
\frac12 (\frac{1}{K_{c,s}} + K_{c,s})$, which are expressed through
the charge and spin velocities $v_{c} \neq v_s$, the charge and spin
stiffnesses $K_{c} \neq K_s$, and the charging energy $E_c$.

In Eq.~\eq{spincond} the rates $\Gamma_l$ and $\Gamma_r$ remain the same
as in the spinless case, since we assume that the density of states in the
leads is spin independent and equals $\nu_{l/r}$ for each spin component.
The spin dependence appears in the functions
$A_{\eta\sigma} = A_{\eta\sigma}^+ + A_{\eta\sigma}^-$,
\beq
A_{\eta \sigma}^{\pm} &=& \sum_{p_c , p_s =0}^{\infty} \frac{g_{p_c}
(\frac12 \gamma_c) g_{p_s}
(\frac12 \gamma_s)}{\omega_c p_c +\omega_s p_s +\bar{a}_0 \pm P_{\eta\sigma}},
\label{alrsig} \\
P_{\eta\sigma} &=&
2 \bar{a}_0 \tilde{N}_{\eta \sigma} + 2 \bar{a}_1 \tilde{N}_{-\eta, \sigma}
\nonumber \\
 &+& 2 \bar{a}_2 \tilde{N}_{\eta, -\sigma}
+ 2 \bar{a}_3 \tilde{N}_{-\eta, -\sigma} .
\eeq
The coefficients $\bar{a}_{0,1} = \frac{\omega_0}{8}
(\tilde{\nu}_c \pm \lambda_c +
\nu_s \pm \lambda_s)$ and $\bar{a}_{2,3} = \frac{\omega_0}{8}
(\tilde{\nu}_c \pm
\lambda_c - \nu_s \mp \lambda_s)$ are the components of the
quadratic form of  the zero-mode Hamiltonian
\beq
H_0 &=& \sum_{\eta,\sigma } \left( \bar{a}_0 \tilde{N}_{\eta\sigma}^2 +
\bar{a}_1 \tilde{N}_{\eta \sigma} \tilde{N}_{-\eta,\sigma} \right.
\nonumber \\
&+& \left. \bar{a}_2 \tilde{N}_{\eta\sigma} \tilde{N}_{\eta, -\sigma}
+ \bar{a}_3 \tilde{N}_{\eta \sigma} \tilde{N}_{-\eta, -\sigma} \right).
\label{nondh0}
\eeq

In the basis
\beq
N_{c,s} &=& (N_{R+} + N_{L+}) + \sigma (N_{R-} + N_{L-}), \label{ncsb}  \\
J_{c,s} &=& (N_{R+} - N_{L+}) + \sigma (N_{R-} - N_{L-}), \label{jcsb}
\eeq
the Hamiltonian \eq{nondh0} becomes diagonal:
\be
H_0 = \frac{\omega_0}{8} \left[\tilde{\nu}_c \tilde{N}_c^2 + \lambda_c
\tilde{J}_c^2 + \nu_s \tilde{N}_s^2 + \lambda_s \tilde{J}_s^2 \right]
\ee
and
\beq
P_{\eta \sigma} = \frac{\omega_0}{4} \left[\left(\tilde{\nu}_c \tilde{N}_c
+\eta \lambda_c \tilde{J}_c \right) + \sigma \left(\nu_s \tilde{N}_s
+\eta \lambda_s \tilde{J}_s \right)\right],
\eeq
where $\tilde{N}_{c,s} = N_{c,s} - 4 k_{\mu,B}$ and $\tilde{J}_{c,s}
= J_{c,s} - 4 k_{\Phi, R}$.

In Eq.~\eq{spincond} the two components $\sigma = \pm$ seem to be
independent of each other. However, this is not the case, and they are,
in fact, entangled due to the nontrivial procedure of averaging over
zero modes.

\section{Averaging over zero modes}

\subsection{Spinless case}

The typical expression to be averaged over zero modes before
the time integration has the form [cf. Eq.~\eq{Gtime}]
\be
\langle e^{i b_1 \tilde{N}+ i b_2 \tilde{J}}\rangle_{z.m.} =
\frac{{\rm Tr} (e^{i b_1 \tilde{N}+ i b_2 \tilde{J}}e^{-\beta H_0})}{{\rm Tr} 
(e^{-\beta H_0})},
\label{avdef0}
\ee
where $b_{1,2}$ depend linearly on the time arguments $t_{1,2}$.
The trace operation is understood as a summation over all possible
integer values of $N_L$ and $N_R$. In the basis $(N, J)$ we have to
sum over either both even $( 2m, 2n )$ or both odd $( 2m+1, 2n+1 )$
eigenvalues. Thus,
\beq
& & {\rm Tr} \left( e^{-\beta H_0} \right) =  \sum_{m,n =-\infty}^{\infty}
e^{-\beta \omega_0 [\tilde{\nu} (m -k_{\mu})^2 + \lambda (n- k_{\Phi})^2]} 
\nonumber
\\
& & \qquad +  \sum_{m,n =-\infty}^{\infty}   e^{-\beta \omega_0 [\tilde{\nu}
(m+\frac12 - k_{\mu})^2 + \lambda (n+ \frac12 - k_{\Phi})^2]}\nonumber
\\ & & = \frac{\pi}{\beta \omega_0 \sqrt{\tilde{\nu} \lambda}}
\left[\theta_3 \left(\pi k_{\mu}, e^{-\frac{\pi^2}{\beta \omega_0
\tilde{\nu}}} \right) \theta_3 \left(\pi k_{\Phi}, e^{-\frac{\pi^2}{\beta 
\omega_0 \lambda}} \right) \right.
\nonumber \\
& & \left. \qquad + \theta_4 \left(\pi k_{\mu}, e^{-\frac{\pi^2}
{\beta \omega_0 \tilde{\nu}}} \right) \theta_4 \left(\pi k_{\Phi},
e^{-\frac{\pi^2}{\beta \omega_0 \lambda}} \right) \right],
\eeq
where $\theta_{3,4}$ are the Jacobian theta functions.
Some properties of these functions are reviewed in Appendix~\ref{jacobi}.

The numerator in Eq. \eq{avdef0} can be equivalently rewritten in the form
\be
{\rm Tr} \left(e^{i b_1 \tilde{N}+ i b_2 \tilde{J}} e^{-\beta H_0} \right) =
{\rm Tr} \left( e^{-\beta H'_0} \right) \cdot e^{-\frac{1}{\beta \omega_0}
\left(\frac{b_1^2}{\tilde{\nu}}+\frac{b_2^2}{\lambda} \right)},
\label{avdef1}
\ee
where $H'_0$ is obtained from $H_0$ by replacing
\beq
k_{\mu} \to k'_{\mu} = k_{\mu} + \frac{i b_1}{\beta \omega_0 \tilde{\nu}}, \\
k_{\Phi} \to k'_{\Phi} = k_{\Phi} + \frac{i b_2}{\beta \omega_0 \lambda}.
\eeq
At low temperatures $\beta^{-1} \ll \omega_0, E_c$,  the last exponential 
factor in Eq. \eq{avdef1} can be discarded. Using Eq. \eq{thrat2} we derive 
the following expression:
\beq
& & \langle e^{i b_1 \tilde{N}+ i b_2 \tilde{J}}\rangle_{z.m.}  \approx
p_1 (k_{\mu}, k_{\Phi}) e^{2 i b_1 f(k_{\mu})+ 2 i b_2 f (k_{\Phi})}
\nonumber \\
& & \qquad + p_2 (k_{\mu}, k_{\Phi}) e^{2 i b_1 f(k_{\mu}+1/2) 
+ 2 i b_2 f (k_{\Phi}+1/2)}.
\label{avdef2}
\eeq
The ``sawtooth'' function
\be
f (x) = \sum_{n=1}^{\infty} \frac{(-1)^n}{\pi n} \sin{2 \pi n x}
\label{saw}
\ee
has the period $1$ and equals $f(x) = -x$ for $x \in (-\frac12, \frac12)$ 
and $f(\pm \frac12) =0$. The functions $p_{1,2} (k_{\mu}, k_{\Phi})$ are 
determined by
\beq
p_1 (k_{\mu}, k_{\Phi}) &=& \frac{1}{1+ \frac{\theta_4 \left(\pi k_{\mu},
e^{-\pi^2/\beta \omega_0 \tilde{\nu}} \right) \theta_4 \left(\pi k_{\Phi},
e^{-\pi^2/\beta \omega_0 \lambda} \right)}{\theta_3 \left(\pi k_{\mu},
e^{-\pi^2/\beta \omega_0 \tilde{\nu}} \right) \theta_3 \left(\pi k_{\Phi},
e^{-\pi^2/\beta \omega_0 \lambda} \right)}} \nonumber \\
&\approx & \frac{1}{1+e^{\beta \omega_0 [\tilde{\nu} g_2 (k_{\mu})+
\lambda g_2 (k_{\Phi})]}},
\eeq
and $p_2 (k_{\mu}, k_{\Phi}) = p_1 (k_{\mu}+\frac12 , k_{\Phi}+\frac12)$.
One can observe that $p_1+p_2 =1$. The function $g_2 (x)$ is introduced
in \eq{gas2}.

Let us consider the limit of zero temperature, or  $\beta \to \infty$.
The expression \eq{avdef2} becomes exact in this limit.
Since $b_1$ and $b_2$ are linear in time, we can perform easily all time
integrations.
Thus, the averaging over zero modes effectively results in replacing
$\tilde{N} \to 2 f (k_{\mu}+ \frac12 \delta_{\mu 1(2)})$ and
$\tilde{J} \to 2 f (k_{\Phi}+ \frac12 \delta_{\Phi 1 (2)})$, where
$\delta_{\mu 1} = \delta_{\Phi 1} =0$ and
$\delta_{\mu 2} = \delta_{\Phi 2} =1$,
which assumes further summation over the different topological
realizations (1 and 2) of the ground state with the weight factors $p_1$ 
and $p_2$. For $\beta^{-1} \ll \omega_0$ the latter approximately equal
\beq
p_1 (k_{\mu}, k_{\Phi}) &=& \Theta (-\tilde{\nu} g_2 (k_{\mu})
- \lambda g_2 (k_{\Phi})), \\
p_2 (k_{\mu}, k_{\Phi}) &=& \Theta (\tilde{\nu} g_2 (k_{\mu})
+ \lambda g_2 (k_{\Phi})),
\eeq
and play the role of projectors which divide the elementary cell
$(k_{\Phi} , k_{\mu} ) \in [-\frac12,\frac12] \times [-\frac12, \frac12]$
into two areas (topological sectors).

Let us analyze such a partition of the elementary cell and consider the
(upper right) quadrant defined by $0<k_{\mu}<\frac12$ and $0<k_{\Phi}<\frac12$.
The function $g_2 (x) = x-1/4$ for $0<x<1/2$, and therefore the border
between the areas of $p_1$ and $p_2$ is given by the equation
\be
\tilde{\nu} k_{\mu} + \lambda k_{\Phi} = \frac{\tilde{\nu}+\lambda}{4}
\label{brdl}.
\ee
For repulsive interactions and, moreover, in the presence of $E_c \neq 0$ 
the relation $\tilde{\nu}/\lambda >1$ is fulfilled.

\begin{figure}[t]
\includegraphics[width=8cm,angle=0]{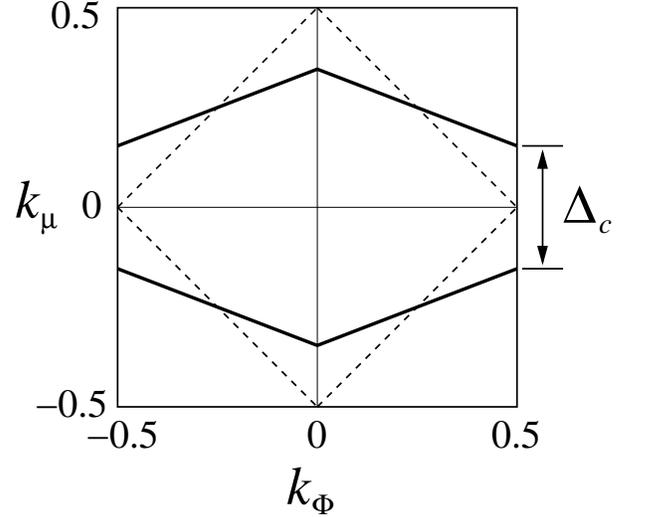}
\caption{Shift of the conductance peaks (solid lines) due
to the charging energy. The dashed lines correspond to the noninteracting case 
($\Delta_c =0$).}
\label{spinlessdiamond}
\end{figure}

We can establish the borders between the topological sectors $p_1$
and $p_2$ in the other quadrants by mirroring Eq. \eq{brdl} with respect
to the $k_{\mu}$ axis, $k_{\Phi}$ axis, or both. Thus, we obtain that the
area of the projector $p_1$ is the inner part of the elementary cell
bounded by the hexagon (see Fig.~\ref{spinlessdiamond}).
Respectively, the outer part is the area of $p_2$.

Let us now analyze the conductance in the upper right quadrant.
In the inner part $p_1$ only the zero harmonic $(p=0)$ of $A_{R}^+$
becomes divergent near the border line \eq{brdl}. In the outer part
$p_2$ the zero harmonic $(p=0)$ of $A_{R}^-$ is divergent near the
border line \eq{brdl}. From both sides of the latter the conductance
behaves like
\be
G \propto \frac{1}{\left[ \tilde{\nu} k_{\mu} + \lambda k_{\Phi}
- \frac{\tilde{\nu}+\lambda}{4}\right]^2}.
\ee
This is an expected result as well as the fact that the pole of the conductance
(condition for the resonant tunneling of an electron) matches with a transition
$p_1 \to p_2$ from one topological sector to another.

In order to identify the positions of the conductance peaks, it appears
sufficient to consider just the zero harmonics $p=0$ of the functions
$A^{\pm}_{\eta}$, because the higher ones ($p \geq 1$) do not have any poles
at all.

It is instructive to derive the conductance in the noninteracting limit
$\tilde{\nu} = \lambda =1$. Using the identity
\be
\frac{\pi}{\sin \pi x} = \sum_{p=-\infty}^{\infty} \frac{(-1)^p}{p+x},
\ee
one can show that
\be
G = \frac{e^2}{2 \pi} \frac{\Gamma_l \Gamma_r}{\omega_0^2}
\left[ \frac{\pi}{\sin (\frac{\Delta \mu}{\omega_0} + k_{\Phi})
\pi}+ \frac{\pi}{\sin (\frac{\Delta \mu}{\omega_0} - k_{\Phi})\pi}\right]^2 .
\label{lowt1}
\ee
This result can be recovered in the scattering matrix approach, if the width
of the Breit-Wigner resonance in \eq{lowt} is neglected. The expression
\eq{lowt1} suggests that the resonant condition can be satisfied at any
$k_{\mu}$ by tuning the magnetic flux $k_{\Phi}$.

As was discussed in Ref.~\onlinecite{kin}, the main qualitative feature imposed
by an electron-electron interaction and/or charging energy is the opening
of a window at certain values of $k_{\mu}$ inside which the resonant
condition is never met. This situation is shown in
Fig.~\ref{spinlessdiamond}, and the corresponding gap
value equals $\Delta_c = \frac12 (1-\frac{\lambda}{\tilde{\nu}})$.

\subsection{Spinful case}

In order to implement an averaging similar to \eq{avdef0} in the
spinful cases, it is  necessary to calculate first the partition
function ${\rm Tr} (e^{-\beta H_0})$. The trace operation is now
understood as a summation over all integer values of $N_{R +}$,
$N_{L +}$, $N_{R -}$, and $N_{L-}$. However, in this basis the
Hamiltonian \eq{nondh0} is not diagonal, and we have to use the
basis \eq{ncsb},\eq{jcsb} instead. The summation rules for the latter have
been formulated, for instance, in Refs.~\onlinecite{faz,mosk,plet}. Applying 
them, one can find that the partition function is proportional to
\beq
\sum_{i=1}^{16} \left[ \theta_3 \left(\pi k_{\mu i}, e^{-\frac{\pi^2}{2 \beta 
\omega_0 \tilde{\nu}_c}}
\right) \theta_3 \left(\pi k_{\Phi i}, e^{-\frac{\pi^2}{2 \beta \omega_0 
\lambda_c}} \right) \right.\nonumber \\
\times \left. \theta_3 \left(\pi k_{B i}, e^{-\frac{\pi^2}{2 \beta \omega_0 
\nu_s}} \right) \theta_3
\left(\pi k_{R i}, e^{-\frac{\pi^2}{2 \beta \omega_0 \lambda_s}} \right) 
\right],
\eeq
where $k_{X i} = k_X +\frac14 \delta_{X i} \,\,  (X=\mu, \Phi, B, R)$
and the summation is performed over 16 topological sectors. The latter are 
specified by $\delta_{X i}$ given in the table
\begin{center}
\begin{tabular}{|c|c|c|c|c|c|c|c|c|c|c|c|c|c|c|c|c|} \hline
$i$ & 1 & 2 & 3 & 4 & 5 & 6 & 7 & 8 & 9 & 10 & 11 & 12 & 13 & 14 & 15 & 16 \\
\hline
$\delta_{\mu i}$  &
0 & 2 & 2 & 0 & 2 & 2 & 0 & 0 & 1 & 3 & 3 & 1 & 3 & 3 & 1 & 1  \\ \hline
$\delta_{\Phi i}$  &
0 & 2 & 2 & 0 & 0 & 0 & 2 & 2 & 1 & 3 & 3 & 1 & 1 & 1 & 3 & 3  \\ \hline
$\delta_{B i}$  &
0 & 2 & 0 & 2 & 2 & 0 & 2 & 0 & 1 & 3 & 1 & 3 & 3 & 1 & 3 & 1  \\ \hline
$\delta_{R i}$  &
0 & 2 & 0 & 2 & 0 & 2 & 0 & 2 & 1 & 3 & 1 & 3 & 1 & 3 & 1 & 3  \\ \hline
\end{tabular}
\end{center}

One can define 16 functions ($i=1,\ldots,16$)
\be
p_i (k_X) = p_1 (k_{X i}),
\label{pi16}
\ee
where $p_1 (k_X)$ equals to
\beq
 & & \frac{1}{1+\sum_{j=2}^{16} \frac{\theta_3 ( \pi k_{\mu j})}{\theta_3
 (\pi k_{\mu})} \frac{\theta_3 (\pi k_{\Phi j})}{\theta_3 (\pi k_{\Phi})}
 \frac{\theta_3 (\pi k_{B j})}{\theta_3 (\pi k_{B})} \frac{\theta_3 
(\pi k_{R j})}{\theta_3 (\pi k_{R})}} \nonumber \\
& & \approx \frac{1}{1 + \sum_{j=2}^{16} e^{2 \beta \omega_0 B_j (k_X)}} ,
\eeq
and the functions $B_j (k_X)$ are introduced in  Appendix \ref{jacobi}.

The functions \eq{pi16} satisfy the identity
\be
\sum_{i=1}^{16} p_i (k_X) =1.
\ee
At low temperatures $\beta^{-1} \ll \omega_0$ we have an approximate relation
\be
p_1 (k_X) = \prod_{j=2}^{16} \Theta (- B_j (k_X)),
\ee
and the functions $p_i (k_X)$ become the projectors which divide the elementary
cell $k_X \in [-\frac12,\frac12]\times \cdots \times [-\frac12,\frac12]$ in the
four-dimensional parameter space into 16 topological sectors.

We can now formulate the rule which prescribes how to evaluate the average over
zero modes in \eq{spincond}: it is necessary to replace
\beq
\tilde{N}_c & \to & 4 f (k_{\mu i}), \quad \tilde{J}_c \to 4 f (k_{\Phi i}), \\
\tilde{N}_s & \to & 4 f (k_{B i}), \quad \tilde{J}_s \to 4 f (k_{R i}),
\eeq
and to sum over $i=1,\ldots,16$ topological realizations of the ground state 
with the weight functions $p_i (k_X)$.

Once this procedure is implemented, it becomes sufficient to
consider just the zero harmonics ($p_c=p_s=0$) of the functions
$A_{\eta \sigma}^{\pm}$, Eq. \eq{alrsig}, for establishing the positions
of the conductance maxima. In this respect there exists a full analogy
with the spinless case, and we refer to the corresponding discussion
in the previous subsection.

\begin{figure}[t]
\includegraphics[width=8.4cm,angle=0]{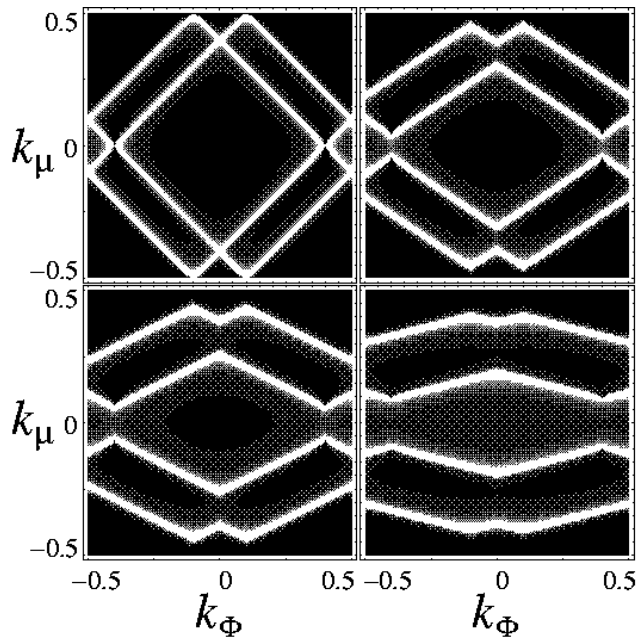}
\caption{Conductance peaks for $k_R=0.1$ and different values of the
charging energy $8 E_c/\omega_0=$ 0.0 (upper left), 0.5 (upper right),
1.0 (bottom left),  3.0 (bottom right). The TLL parameters are 
$\nu_{c,s}=\lambda_{c,s}=1$.}
\label{spinfuldiamond}
\end{figure}

In the framework of the developed formalism it is possible to study
the influence of the TLL interaction on the distribution of the
conductance peaks in the presence of magnetic flux and SO coupling. The
charging effects are  also naturally incorporated, and the charging
energy $E_c$ plays a role similar to that of the TLL parameter
$\nu_c$. They are both combined into $\tilde{\nu}_c$ [see
Eq.~\eq{vcspin}], and therefore the effects produced by each of them
are analogous. Let us then fix $\nu_c =1$ and vary $E_c$. In
Fig.~\ref{spinfuldiamond} we show the elementary cells of the
conductance contour plot in the  $(k_{\Phi}, k_{\mu})$ plane for
$k_R =0.1$ and different values of the charging energy. The TLL
parameters are $\nu_{c,s} = \lambda_{c,s} =1$. One can see how the
separate effects of SO coupling and charging energy (shown in
Figs.~\ref{condpeakLB} and \ref{spinlessdiamond}, respectively)
merge together.

In experiments the usual tuning parameters are $\Phi$ and $V_g$. The
parameter $V_g$  appears in the theoretical model through both $k_R$
and $k_{\mu}$. The Rashba coupling constant depends on an applied
gate voltage\cite{nitta}, which is modeled  by  $\alpha_R =
\alpha_R^0 - \frac{\pi}{m^*L} \kappa v_g$, where $v_g = \frac{e
V_g}{\omega_0}$ and $\kappa>0$ is a dimensionless  coefficient. The
degeneracy point in gate voltage at which  the Rashba coupling
$\alpha_R$ vanishes is defined by $v_g^0 \equiv v_g (\alpha_R=0) =
\frac{m^* L \alpha_R^0}{\kappa \pi}$. Introducing  the departure
from the degeneracy point $\Delta v_g = v_g -v_g^0$, we then express
\be
k_R (\Delta v_g) = \frac12 \left[ \sqrt{1+ \kappa^2 (\Delta v_g)^2} -1 \right]
\ee
and
\be
k_{\mu} (\Delta v_g) = \frac{\Delta v_g + v'}{1+8 E_c/\omega_0}.
\ee
Here $v' =  v_g^0 + \frac{\Delta \mu}{\omega_0} + N_0 - \frac12$
determines the shift of the whole pattern; we may put at will $v'=0$.

\begin{figure}[t]
\includegraphics[width=8.4cm,angle=0]{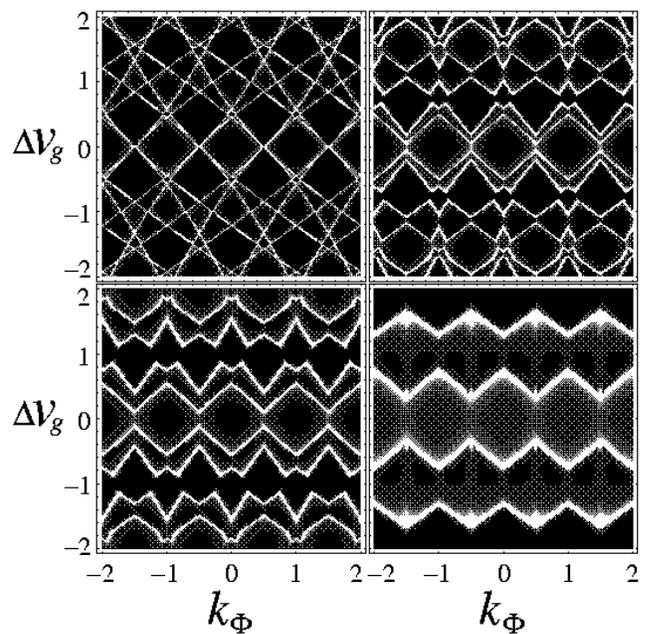}
\caption{Conductance peaks for $\kappa=1$ and different values of the charging
energy $8 E_c/\omega_0 =$ 0.0 (upper left),  0.5 (upper right), 1.0
(bottom left), 3.0 (bottom right).}
\label{gvdep}
\end{figure}

In Fig.~\ref{gvdep} we demonstrate the influence of the charging effects
in $(k_{\Phi}, \Delta v_g)$ plane. The values of $E_c$ and $\nu_{c,s}$,
$\lambda_{c,s}$ are the same as in Fig.~\ref{spinfuldiamond}. We observe
that upon enhancement of the charging energy the gap opens near
$\Delta v_g =0$. Due to the presence of the gate-dependent SO coupling,
the pattern of conductance maxima is more complicated than that discussed
in Ref.~\onlinecite{kin}.

\section{Discussion and conclusion}
In this work we have studied the tunneling conductance of a
mesoscopic one-dimensional ring attached to two Fermi reservoirs.
The interaction inside the ring is described by the
Tomonaga-Lutinger liquid. The bosonization approach which is usually
adopted for the study of such model allows us to include the flux and
gate-voltage dependence as well as the influence of SO coupling. It
is remarkable that all these externally tuned parameters appear in
the topological sector of the bosonized theory. Therefore, the
accurate treatment  of zero modes is required in order to describe
the mesoscopic phenomena at low temperatures $\beta^{-1} \ll
\omega_0$.

Using the Keldysh formalism we have performed the calculation of a
linear conductance in the limit of weak tunneling between the leads
and the ring. The real-time approach  allows one to obtain asymptotic
results for the distribution of the conductance peaks in the low
temperature limit. Although the perturbative expansion of the
conductance in the tunneling strength is not very well suited for a
description of a shape of conductance peaks, it is nevertheless
quite efficient for establishing their positions. The topological
origin of the peaks' distribution alludes to its robustness
upon small modifications of the model.

We have studied the patterns of the conductance maxima at nonzero
spin-orbit coupling as a function of magnetic flux and gate voltage.
The Tomonaga-Luttinger interaction and the charging energy have been
seen to contribute in analogous way. In rings of reduced size
the account of charging effects might appear more experimentally
motivated, and therefore we concentrated on their study. In
particular, we have made a theoretical prediction for the
distribution of the conductance peaks in the presence of both the
charging energy and the spin-orbit coupling. We observed an
interesting interplay between both effects. The SO coupling
lifts the degeneracy of the conduction peaks, and the charging
energy opens a gap centered at the remaining points of the
degeneracy in question. The value of this gap is proportional to the
charging energy. When the latter becomes very large, the Rashba
effect is less pronounced. The pattern of the conductance peaks then
approaches the form of hexagonal honeycombs which is typical to
spinless fermions.

In conclusion, we have described the interplay between Coulomb
blockade and Aharonov-Bohm and Aharonov-Casher effects for the
different values of the charging energy, magnetic flux, and
spin-orbit coupling, as is manifested in the contour plots of the
tunneling conductance.

\begin{acknowledgments}
We would like to thank Dmitry Bagrets, Thierry Champel, Ilya Krive, and Gerd
Sch{\"o}n for fruitful discussions. M.P. was supported by the DFG
Center for Functional Nanostructures at the University of Karlsruhe.
V.G. was supported by the Swiss National Science Foundation.
\end{acknowledgments}

\appendix

\section{Correlation functions of the Tomonaga-Luttinger model}
\label{bosonic}

Let us consider for simplicity the spinless case. The canonical transformation
which solves the two-channel TLL model is
\be
\left( \begin{array}{c} d_{1k} \\ d^{\dagger}_{2k} \end{array}\right) =
\left( \begin{array}{cc} u_+ & u_- \\  u_- & u_+ \end{array}\right)
\left( \begin{array}{c} b_{Rk} \\ b^{\dagger}_{Lk} \end{array}\right),
\label{bogtr}
\ee
where $u_{\pm} = \sqrt{(\gamma \pm 1)/2}$.

The explicit form of the time evolution of the bosonized fields reads
\beq
& & \psi_{\eta}^b (x,t)= \frac{1}{\sqrt{L \tilde{\alpha}}} e^{- i \sqrt{2 \pi} 
\phi_{\eta} (x,t)}, \\
& & \phi_{\eta} (x,t)= i \sum_{k=1}^{\infty}
\frac{e^{-\frac12 \tilde{\alpha} k}}{\sqrt{2 \pi k}} \nonumber \\
& & \times \left[ u_{\eta} D_{1k} (x-\omega t) - u_{-\eta} D_{2k} (x+\omega t)
\right],
\eeq
where
\beq
D_{1k} &=& e^{i k x} d_{1k} - e^{-i k x} d_{1k}^{\dagger}, \\
D_{2k} &=& e^{i k x} d_{2k}^{\dagger} - e^{-i k x} d_{2k}.
\eeq

Let us consider the correlation function
\beq
\langle \psi_{\eta}^b (x,t) \psi_{\eta}^{b\dagger} \rangle &=& e^{-\pi\langle
(\phi_{\eta} (x,t) -\phi_{\eta})^2\rangle+\pi [\phi_{\eta} (x,t), \phi_{\eta}]}
\nonumber \\
&=& e^{-u_+^2 (\pi {\cal D} (\eta x-\omega t)- i \chi ( \eta x-\omega t))}
\nonumber \\
&\times & e^{-u_-^2 (\pi {\cal D} (\eta x+\omega t)
+ i \chi (\eta x +\omega t))},
\label{corrfun}
\eeq
where
\beq
{\cal D} (x) &=& \frac{1}{\pi} \sum_{k=1}^{\infty} \frac{1 -\cos k x}{k}
\left(\frac{2}{e^{\beta \omega k}-1} + e^{-\tilde{\alpha} k} \right), \\
\chi (x) &=& \sum_{k=1}^{\infty} \frac{\sin k x}{k} e^{-\tilde{\alpha} k} =
\frac{1}{2 i} \ln \frac{1-e^{- i x -\tilde{\alpha}}}{1-e^{i x -\tilde{\alpha}}}
\eeq
are the periodic functions of $x$. Obviously,  the function \eq{corrfun}
is also periodic in real time, and therefore it can be expanded in a Fourier 
series with the frequency $\omega$.

In the zero-temperature limit $\beta \to \infty$ the temperature-dependent
part of ${\cal D} (x)$ can be discarded, and we obtain
\be
{\cal D} (x) = - \frac{1}{2 \pi} \ln \frac{(1 -e^{-\tilde{\alpha}})^2}
{(1-e^{- i x -\tilde{\alpha}}) (1-e^{ i x -\tilde{\alpha}})}.
\ee
Hence, the correlation function \eq{corrfun} is equal to
\be
\left( \frac{1 -e^{-\tilde{\alpha}}}{1-e^{i (\eta x - \omega t 
+ i \tilde{\alpha})}}\right)^{u_+^2}
\left( \frac{1 -e^{-\tilde{\alpha}}}{1-e^{-i (\eta x + \omega t 
- i \tilde{\alpha})}}\right)^{u_-^2}.
\ee

For $x =\pi$ we have
\be
\langle \psi_{\eta}^b (\pi,t) \psi_{\eta}^{b\dagger} \rangle =
\frac{1}{L \tilde{\alpha}} \left( \frac{1 -e^{-\tilde{\alpha}}}
{1+e^{-i \omega t - \tilde{\alpha}}}\right)^{\gamma} .
\ee
Its Fourier expansion \eq{fourex} is equivalent to the Taylor
expansion of the analytic function $(1+z)^{-\gamma}$ for $|z|<1$.
Therefore the expansion \eq{fourex} contains only non-negative Fourier
harmonics ($p \geq 0$) with the coefficients
\beq
g_p (\gamma) &=& \frac{(-1)^p \Gamma (p+\gamma)}{\Gamma (\gamma) \Gamma (p+1)}
\frac{(1-e^{-\tilde{\alpha}})^{\gamma}}{L \tilde{\alpha}} 
e^{-p \tilde{\alpha}}\nonumber \\
&\approx & \frac{(-1)^p \Gamma (p+\gamma)}{\Gamma (\gamma) \Gamma (p+1)} 
\frac{\tilde{\alpha}^{\gamma-1}}{L} .
\label{fourcoef}
\eeq

Computation of a correlation function in the spinful case is analogous.
A new feature arising in this case is the double time periodicity  
of correlation functions with frequencies $\omega_c$ and
$\omega_s$, which are in general incommensurate.

\section{Jacobian $\Theta$ functions}
\label{jacobi}

The Jacobian function $\theta_3$ is defined by
\be
\theta_3 (z, q) = 1+ 2 \sum_{n=1}^{\infty} q^{n^2} \cos 2 n z,
\ee
and the  Jacobian function $\theta_4$ can be expressed as
\be
\theta_4 (z, q) = \theta_3 \left( z+\frac{\pi}{2}, q\right).
\ee
Both functions are periodic under the shift $z \to z+\pi$.

Making a Poisson summation, one can prove that
\be
\sum_{k=-\infty}^{\infty} e^{-a (k+z)^2} =
\sqrt{\frac{\pi}{a}} \theta_3 \left( \pi z, e^{-\pi^2/a} \right).
\ee

Using the expression for the ratio of two $\theta_3$ functions,
\be
\ln \frac{\theta_3 (z_1+z_2, q)}{\theta_3 (z_1 - z_2, q)} =
4 \sum_{n=1}^{\infty} \frac{(-1)^n}{n} \frac{q^n}{1- q^{2n}} \sin 2 n z_1 
\sin 2 n z_2,
\label{thrat1}
\ee
one can establish that
\be
\lim_{\beta \to \infty} \frac{\theta_3 \left(\pi (x+\frac{i b}{\beta \omega_0 
\lambda}),
e^{-\frac{\pi^2}{2 \beta \omega_0 \lambda}} \right)}{\theta_3 
\left(\pi x, e^{-\frac{\pi^2}
{2 \beta \omega_0 \lambda}}\right)} = e^{4 i b f(x)},
\label{thrat2}
\ee
where $f (x)$ is a sawtooth function introduced in \eq{saw},
as well as that
\be
\lim_{\beta \to \infty} \frac{\theta_3 \left(\pi (x+\frac{m_1}{4}),
e^{-\frac{\pi^2}{2 \beta \omega_0 \lambda}} \right)}
{\theta_3 \left(\pi (x+\frac{m_2}{4}), e^{-\frac{\pi^2}{2 \beta \omega_0 
\lambda}}\right)}
\approx e^{2 \beta \omega_0 \lambda g_{m_1 m_2} (x)},
\label{thrat3}
\ee
where $m_1,m_2=0,1,2,3$ and
\beq
& & g_{m_1 m_2} (x) = \sum_{n=1}^{\infty} \frac{(-1)^n}{\pi^2 n^2}
\nonumber \\
& & \times  \left[\cos \frac{\pi n}{2} (4 x + m_2) 
- \cos \frac{\pi n}{2} (4 x+m_1) \right].
\label{gfour}
\eeq

In the Fourier series \eq{gfour} one can recognize the functions
\beq
g_{1} (x) & \equiv & g_{10} (x)=-g_{01} (x) \label{gas1} \\
&=&  \left| \left\{\frac12 +x\right\} - \frac34 \right| 
+ \frac12 \left\{\frac12 +x \right\} - \frac{9}{16}, \nonumber \\
g_{2} (x)  & \equiv & g_{20} (x)=-g_{02} (x) \label{gas2} \\
&=&  - \left| \left\{ x \right\}- \frac12 \right| + \frac14, \nonumber\\
g_{3} (x) & \equiv &  g_{30} (x)=-g_{03} (x) \label{gas3} \\
&=& \left| \left\{\frac12 - x\right\} - \frac34 \right| 
+ \frac12 \left\{\frac12 -x \right\} - \frac{9}{16}, \nonumber
\eeq
where $\{x\} \equiv (x \, {\rm mod} \, 1)$ is a fractional part of $x$. 
The function $\{x\}$ has a period $1$ and possesses a property 
$\{- x\} = 1- \{x\}$. One can notice that $g_3 (x) = g_1 (-x)$.

The other functions $g_{m_1 m_2}$ are also expressed in terms of 
$g_1$, $g_2$, $g_3$:
\beq
g_{12} (x)=-g_{21} (x) = g_3 (x-1/2), \\
g_{32} (x)=-g_{23} (x) = g_1 (x+1/2), \\
g_{31} (x)=-g_{13} (x) = g_2 (x+1/4).
\eeq

We also define the following functions:
\beq
B_2 &=& \tilde{\nu}_c g_2 (k_{\mu}) +\lambda_c g_2 (k_{\Phi}) 
+ \nu_s g_2 (k_B)+ \lambda_s g_2 (k_R), \nonumber \\
B_3 &=& \tilde{\nu}_c g_2 (k_{\mu})+ \lambda_c g_2 (k_{\Phi}) , \nonumber \\
B_4 &=& \nu_s g_2 (k_B) +\lambda_s g_2 (k_R), \nonumber \\
B_5 &=& \tilde{\nu}_c g_2 (k_{\mu}) + \nu_s g_2 (k_B), \nonumber \\
B_6 &=&  \tilde{\nu}_c g_2 (k_{\mu}) + \lambda_s g_2 (k_R), \nonumber \\
B_7 &=& \lambda_c g_2 (k_{\Phi}) + \nu_s g_2 (k_B), \nonumber \\
B_8 &=& \lambda_c g_2 (k_{\Phi}) + \lambda_s g_2 (k_R), \nonumber \\
B_9 &=& \tilde{\nu}_c g_1 (k_{\mu}) + \lambda_c g_1 (k_{\Phi})  
+  \nu_s g_1 (k_B) + \lambda_s g_1 (k_R), \nonumber \\
B_{10} &=&  \tilde{\nu}_c g_3 (k_{\mu}) + \lambda_c g_3 (k_{\Phi})  
+ \nu_s g_3 (k_B) + \lambda_s g_3 (k_R), \nonumber \\
B_{11} &=& \tilde{\nu}_c g_3 (k_{\mu}) + \lambda_c g_3 (k_{\Phi}) 
+  \nu_s g_1 (k_B) + \lambda_s g_1 (k_R), \nonumber \\
B_{12} &=& \tilde{\nu}_c g_1 (k_{\mu}) + \lambda_c g_1 (k_{\Phi})  
+  \nu_s g_3 (k_B)+ \lambda_s g_3 (k_R), \nonumber \\
B_{13} &=& \tilde{\nu}_c g_3 (k_{\mu}) + \lambda_c g_1 (k_{\Phi})  
+  \nu_s g_3 (k_B)+ \lambda_s g_1 (k_R), \nonumber \\
B_{14} &=& \tilde{\nu}_c g_3 (k_{\mu}) + \lambda_c g_1 (k_{\Phi})  
+  \nu_s g_1 (k_B)+ \lambda_s g_3 (k_R), \nonumber \\
B_{15} &=& \tilde{\nu}_c g_1 (k_{\mu}) + \lambda_c g_3 (k_{\Phi})  
+  \nu_s g_3 (k_B)+ \lambda_s g_1 (k_R), \nonumber \\
B_{16} &=& \tilde{\nu}_c g_1 (k_{\mu})  + \lambda_c g_3 (k_{\Phi})  
+  \nu_s g_1 (k_B) + \lambda_s g_3 (k_R). \nonumber
\eeq

\end{document}